\pdfoutput=1
\documentclass[preprint,12pt,3p]{elsarticle}

\usepackage{amssymb}
\usepackage[table]{xcolor}
\usepackage{booktabs}
\usepackage{amsmath}

\journal{Mechanics of Materials}

\begin{document}

\begin{frontmatter}

\title{Manufacturing of 3D-metallic electromagnetic metamaterials for feedhorns used in radioastronomy and satellite communications}

\author[label1,label2]{Javier De Miguel-Hern\'andez\corref{cor1}}
\address[label1]{Instituto de Astrof\'isica de Canarias, E-38200 La Laguna, Tenerife, Spain\\}
\address[label2]{Departamento de Astrof\'isica, Universidad de La Laguna, E-38206 La Laguna, Tenerife, Spain}

\cortext[cor1]{Corresponding author}

\ead{jmiguel@iac.es}
%\ead[url]{iac.es}

\author[label1,label2]{Roger J. Hoyland}
%\ead{rjh@iac.es}

\author[label3]{Dar\'io Sosa-Cabrera}
\address[label3]{Subsea Mechatronics, S.L.}
%\ead{dario.sc@subseamechatronics.com}

\author[label3]{Sebastiaan Deviaene}

\author[label1,label2]{Pablo A. Fuerte-Rodr\'iguez}
\author[label1,label2]{Eduardo D. Gonz\'alez-Carretero}
\author[label1,label2]{Afrodisio Vega-Moreno}

\begin{abstract}
The electromagnetic metamaterials at microwaves frequencies are well established in industrial applications nowadays. Recent research has shown that a specific kind of metallic metamaterial can contribute to improve the performance of the microwave feedhorns used in radioastronomy and satellite telecommunications. In this article, we theoretically justify this argument finding a new type of meta-ring with a record bandwidth in terms of cross-polarization, and we explore the manufacturability of these particular metamaterials, successfully fabricating a meta-ring and applying it to a novel and very compact prototype microwave antenna which covers a 2:1 bandwidth.
\end{abstract}

\begin{keyword}
%% keywords here, in the form: keyword \sep keyword
Metamaterials \sep microwaves\sep  radioastronomy\sep polarimetry \sep antennas \sep satellite
%% MSC codes here, in the form: \MSC code \sep code
%% or \MSC[2008] code \sep code (2000 is the default)
\end{keyword}

\end{frontmatter}

%%
%% Start line numbering here if you want
%%
% \linenumbers

%% main text
\section{Introduction}
\label{Intro}

Electromagnetic metamaterials can be defined as materials with electromagnetic properties given by their geometrical structure which differs from that given by their basic constituent properties. We can for example design a metallic surface with an effective refractive index different from the refractive index of the metal itself giving more desirable properties. 
Metamaterials have been successfully used in order to create invisibility cloaks at microwave frequencies\cite{Schurig} as well, bringing the "dream" of invisibility a step nearer. The use of electromagnetic metamaterials is today well established in industry. Miniaturized antennas commonly designed for mobile telecommunications applications (e.g., \cite{Sievenpiper}, \cite{Sievenpiper2}, \cite{Sievenpiper3}) are a typical instance. 

Recent research has led us to believe that electromagnetic metamaterials can also be used in the manufacture of antennas for satellite communications \cite{Lier}, and our ongoing research indicates that they can also improve the performance of astronomical radio-telescopes by providing ultra wide-band meta-antennas. This feature will bring about a revolution in the state of the art in the field of feedhorn-antennas with low cross-polarization and sidelobes levels, with has been static for several decades due to insurmountable limitations in the bandwidth of the corrugated feedhorns \cite{Clarricoats}. This characteristic is explained in more detail in section \ref{Theory}.

However, given the geometric complexity and the small size of the metamaterials needed for  applications in feedhorns for satellite communications and astronomy, where the lengths, and especially the thicknesses of the three-dimensional metallic pieces are of the order of a fraction of the wavelength of the usable frequency\footnote{The usable frequency is the centre frequency of the transmission/reception band.}, their fabrication is very challenging.

In section \ref{Methods}, we present a mushroom type electromagnetic metamaterial \cite{ADA} manufactured by three different methods, i.e., metallic additive 3D-printing, stereolithography (SLA) 3D-prototyping and a simplified machinable version\footnote{Surprisingly finding a notable electromagnetic benefit, explained in section \ref{Theory}.}, and we perform a metrological study in order to verify that the fabricated prototypes achieve the mechanical tolerances we need for their application in radioastronomy. This tolerance is of the order of magnitude of $\lambda/100$, which is, around 100[$\mu$m] for microwave frequencies of tens of gigahertz.

Since the feedhorns do not do mechanical work, a stress test is not applicable. However, since the feedhorns may well be cooled in a cryostat to 4[K], a theoretical calculation of the differential contraction of the SLA 3D-printing prototype and its copper skin is relevant and is taken on in section \ref{Theory}.

In section \ref{Methods}, various fabrication techniques are described. In section \ref{Results}, we present a metrological study of the manufactured prototypes, while general conclusions are presented in section \ref{Conclusions}.

Success in this research could have important benefits for astronomical applications, because the sensitivity of broadband receivers scales as a factor $\sqrt{\Delta f }$, i.e., the square root of bandwidth obtained in comparison with a classic corrugated horn-antenna. Also this feature could imply a reduction in the complexity of a radiometer design if the same feedhorn could be used to cover what has been two bands up until now. In terms of sensitivity for our telescopes in operation and future radiotelescopes placed at the Observatorio del Teide (OT) \cite{Quijote}, we estimate that this benefit could be as much as 41\% overall sensitivity. Furthermore, a successful manufactured design of meta-horn will also permit a reduction in the number of horn-antennas per experiment for a given frequency band, leading to a reduction in the complexity of the instruments and saving both mechanical and electronic resources.

\section{Theory and calculations}
\label{Theory}
\subsection{Surface impedance calculations for periodic structures}
Radiotelescopes generally use parabolic reflectors. The field generated at the optical focus of a paraboloid with large $f/D$ ratio is composed of  two polarizations \cite{Minnet}. One P-polarization or also called transverse magnetic (TM) component is polarized in the plane of incidence and has its electric (E) component oriented normally to the focal plane and an S-polarization or transverse electric (TE) pattern, polarized normally to the plane of incidence. In this reference, the authors explain that the feedhorn which transmits both TE and TM polarization with null cross-polarization must have inner surfaces satisfying the following expression

\begin{equation}
Z^{S}Z^{P}=-Z_0^2 \;,
\label{equation_1}
\end{equation}

where $Z^{S,P}$ is the surface impedance for both polarizations  (or spatial directions from the surface) and $Z_0$ is the impedance of the vacuum. The relation in (\ref{equation_1}) is known as the {\it balanced hybrid-mode condition} and guaranties that the $HE_{11}$ hybrid mode is the lower cut-off frequency mode, that is, the \textit{fundamental} mode, and of a high cross polarization purity. This is a necessary but not a sufficient condition to attenuate cross-polarization to desirable levels. Other factors such as the profile of the antenna or the antenna-reflector array optics also influence the final levels of cross polarization. 
In this work by Minnet\&Thomas it is even suggested that a radially corrugate cylindrical waveguide be used in order to support the hybrid-mode condition over a limited band. Other authors gave a more in-depth treatment of this corrugated feed-horn technology \cite{Clarricoats}, and that is the most extensive technology used in radioastronomy and satellite communications for the design of horn antennas up until now, presenting a limiting bandwidth factor of around 1.4:1. This bandwidth can be slightly increased by the use of ring-loaded slots in the throat up umtil a limit of 2:1 bandwidth factor, but the cross-polarization behaviour doesn't satisfy the standard requirements in radioastronomy or satellite communications over the entire band (e.g., return-loss better than -20[dB], side-lobes level better than -30[dB], cross-polarization level better than -35[dB] and maximum gain around 20[dB]).

In order to avoid these geometrical limitations  for a bandwidth factor over 2:1 with low cross-polarization levels over the entire band, i.e., satisfying the hybrid-condition, the use of novel metamaterials is key.  

Thus, it is necessary to be able to calculate surface impedance to design new metamaterials with $Z_{S,P}$ taking on the critical value. This is not trivial. We have developed a similar theoretical approach to that presented in \cite{Lier} to correctly run calculations and simulations. This idea is summarized in the following lines.

\begin{figure}[ht!]
		\includegraphics[width=0.5\textwidth]{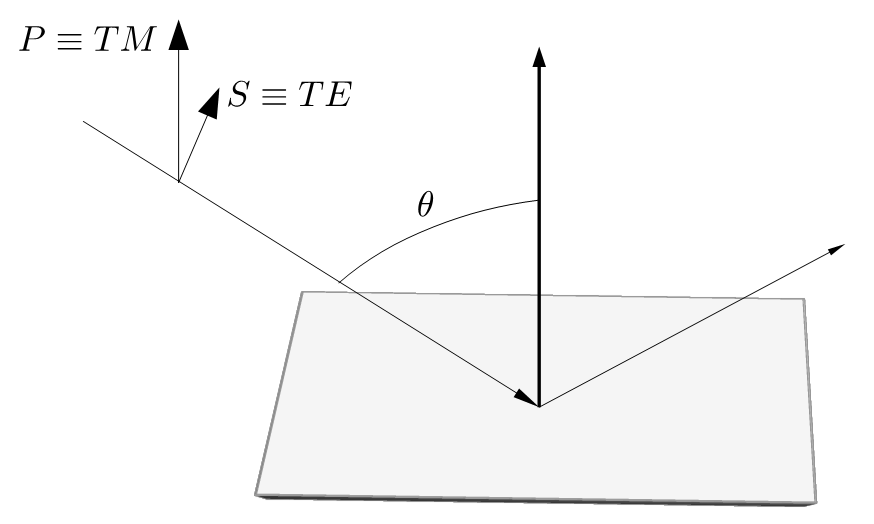}
		\centering 
		\caption{Schematic of S and P polarization of electromagnetic modes}
		\label{fig3_1}
	\end{figure}

It is well known that the reflection coefficient for a wave travelling through vacuum and perpendicularly to a plane surface is given by

\begin{equation}
\Gamma=\frac{E^-}{E^+}=\frac{Z_s-Z_0}{Z_s+Z_0} \;,
\label{equation_2}
\end{equation}

where $E^+$ represents the forward incident wave and $E^-$ represents the backward reflected wave, $Z_s$ is the impedance of the plane surface and $Z_0$ is the impedance of the free space. This is represented in Fig. \ref{fig3_1}. 

Thus, it is possible to obtain expressions for TE and TM modes being reflected by a surface with characteristic $Z^{S,P}$ impedance. In order to obtain these expressions, it must be noted that the impedance of free space ''depends'' on the incidence angle ($\theta$). The expressions are, for P-polarized modes

\begin{equation}
TM\equiv P
\begin{cases} 
 |H\big|= H_0 \\
 |E\big|=E_0\mathrm{cos}\theta
\end{cases} 
\longrightarrow Z_0^{P}(\theta)=\frac{E^{P}(\theta)}{H^{P}(\theta)}=Z_0\mathrm{cos}\theta,
\label{equation_3}
\end{equation}

and for S-polarized modes

\begin{equation}
TE\equiv S
\begin{cases} 
 |H\big|= H_0 \mathrm{cos}\theta\\
 |E\big|=E_0
\end{cases} 
\longrightarrow Z_0^{S}(\theta)=\frac{E^{S}(\theta)}{H^{S}(\theta)}=\frac{Z_0}{\mathrm{cos}\theta}.
\label{equation_4}
\end{equation}

From equations (\ref{equation_2}), (\ref{equation_3}) and (\ref{equation_4}) it is straight forward to obtain the following expression for the surface impedance of reflected S-modes

\begin{equation}
Z_s^{S}=\frac{Z_0}{\mathrm{cos} \theta} \frac{1+\Gamma^{S}}{1-\Gamma^{S}}\;,
\label{equation_5}
\end{equation}

and for reflected P-modes

\begin{equation}
Z_s^{P}=Z_0 \, \mathrm{cos} \theta \, \frac{1+\Gamma^{P}}{1-\Gamma^{P}}\;.
\label{equation_6}
\end{equation}

Thus, we are able to obtain calculations of the surface impedance once we have calculated (from simulation) the reflection coefficient $\Gamma$ of each mode, and then introduced both reflection coefficients in (\ref{equation_1}) in order to know if our metasurface satisfies the hybrid-mode condition. This procedure can be run by using CST Studio Suite$^{\textregistered}$. CST includes a finite element method (FEM) 3D Electromagnetic analysis software package for components and systems. In our simulations, a plane wave is reflected on a metamaterial containing an E-field probe over its surface. Comparing measurements from the E-field probe in the same position with and without the presence of the metamaterial under test, it is possible to subtract the incoming and the reflected signals, allowing us to obtain the reflection coefficient for each S or P polarization from the equations (\ref{equation_5}) and (\ref{equation_6}).

\begin{figure}[ht!]
		\includegraphics[width=1\textwidth]{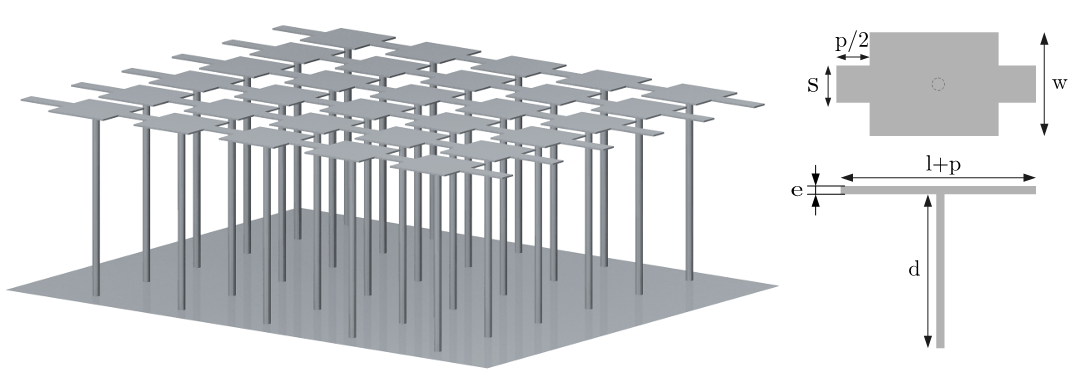}
		\centering 
		\caption{Mushroom type metamaterial based on \cite{ADA}. For covering the band 10-20[GHz], the parameters can adopt the values w=2, s=0.4, p=1, l=2 and d=5.2[mm]. The thickness (e) is 0.05 and the diameter of the cylinders is 0.2[mm]. The lattice period or space between consecutive lines is 0.4[mm].}
		\label{fig3_2}
	\end{figure}

\begin{figure}[ht!]
		\includegraphics[width=0.4\textwidth]{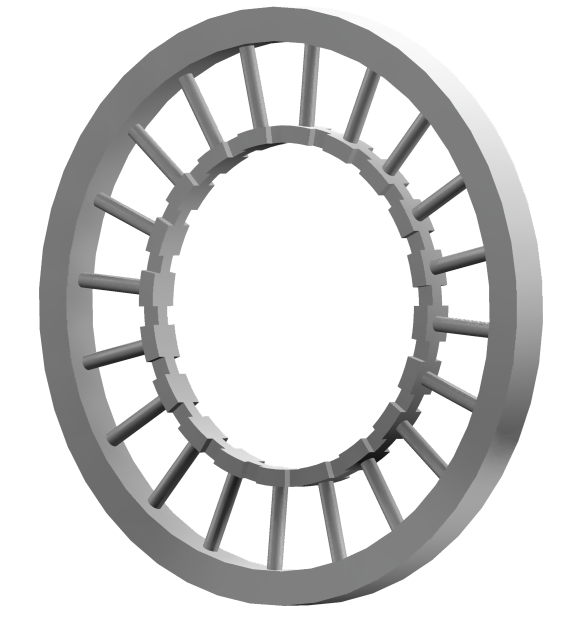}
		\centering 
		\caption{Ring of the meta-horn antenna. Here the more realistic dimensions of the model-2 are used in a ring of 17.4[mm] exterior radius.}
		\label{fig3_3}
	\end{figure}

\begin{figure}[ht!]
		\includegraphics[width=0.8\textwidth]{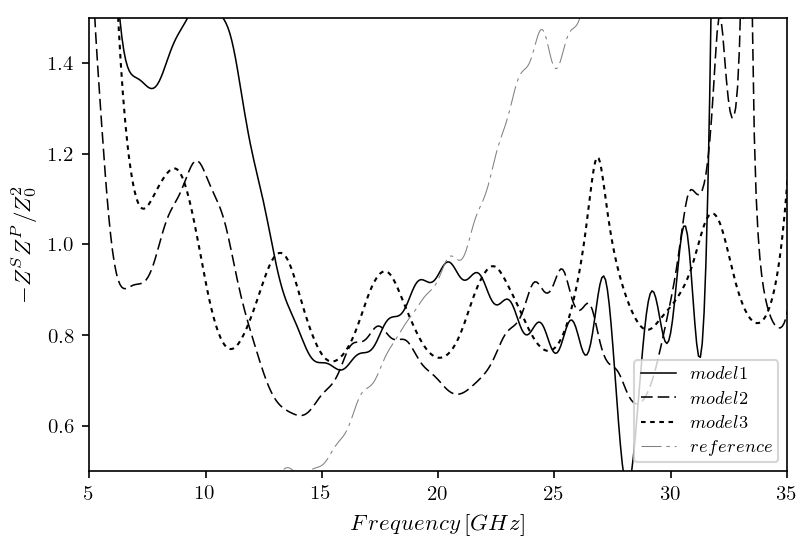}
		\centering 
		\caption{Metamaterial FEM-simulations for 3 different models. The metamaterial satisfies the hybrid mode condition when the expression in the y-axis is exactly 1. The solid line represents the results for the model-1 mushroom topology metamaterial inspired on \cite{ADA}. The dashed line corresponds to the modified  model-1 (model-2) mushroom metamaterial, with cylinder diameter = 0.85, s=0.9 and plate thickness  equal 1[mm]. The dotted line represents both top-plated and inverted-``L'' topologies of model-3, since they present identical results. The dot-stripe gray line represents an equivalent model using corrugations instead of mushrooms to provide a reference. The material is a perfect electric conductor (PEC) for all simulations.}
		\label{fig3_4}
	\end{figure}

These simulations have been taken for a mushroom 3D-type metamaterial based on \cite{ADA} and presented in Fig.\ref{fig3_2}, obtaining the results shown in the Fig.\ref{fig3_4}.  This metamaterial is named as \textit{model-1}. It is convenient to keep in mind now that the aim of this simulations is to design the inner surface of a horn-antenna, made up of tens or hundreds of rings such as the shown in the Fig. \ref{fig3_3} joined together.

From the hybrid-mode condition in (\ref{equation_1}),  a metamaterial satisfies the hybrid condition shown in Fig.\ref{fig3_4}  when $-Z^{S} Z^{P}/Z_0^2$ is exactly 1. Thus, a metamaterial with values $-Z^{S} Z^{P}/Z_0^2=1$ would theoretically allow us to design inner walls for feedhorn antennas with null cross-polarization levels, while it is expected that values of $-Z^{S} Z^{P}/Z_0^2$ in the vicinity of 1 yields  relative cross-polarization attenuation. This is the case of this metamaterial, whose dimensions are given in Fig.\ref{fig3_2} and whose simulation is represented by the solid black line of the Fig.\ref{fig3_4}, showing good behaviour in the band 12-28[GHz] approximately, which is a 2.5:1 bandwidth factor.

However, the small dimensions for the sample of metamaterial in the Fig.\ref{fig3_2} are not possible to achieve in a mechanical workshop or by additive techniques. Thus, for the prototype presented later in the section \ref{Methods} we imposed several constraints. Firstly, since 0.05[mm] thickness for the plates is far from a realistic value a conservative value for this thickness was established as 1[mm]. The diameter of the cylinder is challenging to manufacture and it has been set at 0.85[mm]. Thirdly, the dimension of the parameter "s" had been increased to 0.9[mm]. The rest of parameters of model-2 have the same value as model-1. The effect of these modifications over the model-1 design with the original dimensions based on \cite{ADA} is studied with the help of the results of a new simulation shown in Fig.\ref{fig3_4}, where  this typology is named as \textit{model-2}. It is possible to see how the effect is not significant straying more from the value $-Z^{S} Z^{P}/Z_0^2\sim1$ for centre frequencies but slightly increasing the overall bandwidth to lower frequencies. The explanation for this increment of bandwidth maybe due to the capacitive impedance given by the increase in cross section of the plates and cylinders whilst simultaneously reducing the inductance.

\begin{figure}[ht!]
		\includegraphics[width=0.75\textwidth]{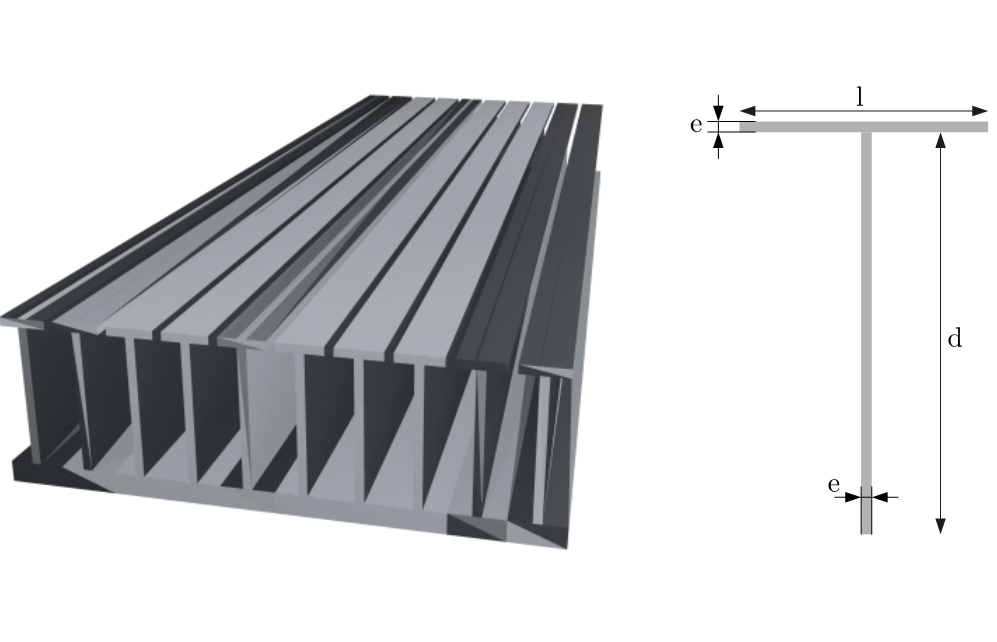}
		\centering 
		\caption{Top-plated corrugations in a simplified version of the original metamaterial where e=0.4, l=1.66 d=6.5[mm] and the free space between corrugations is 0.25[mm]. Note that a classic mechanization permits a reducible thickness (e).}
		\label{fig3_5}
	\end{figure}

We have compared the highly-complex topology of metamaterials based on \cite{ADA} (model-1) and its realistically manufacturable version (model-2) with a simpler version redesigned to be manufactured by ''traditional'' workshop techniques. Now, the cylindrical pattern is reduced to a solid wall, while the plates are also unified, forming a "tee" from a frontal view. This design is presented in Fig.\ref{fig3_5} and named as \textit{top-plated corrugations}. 

\begin{figure}[ht!]
		\includegraphics[width=0.75\textwidth]{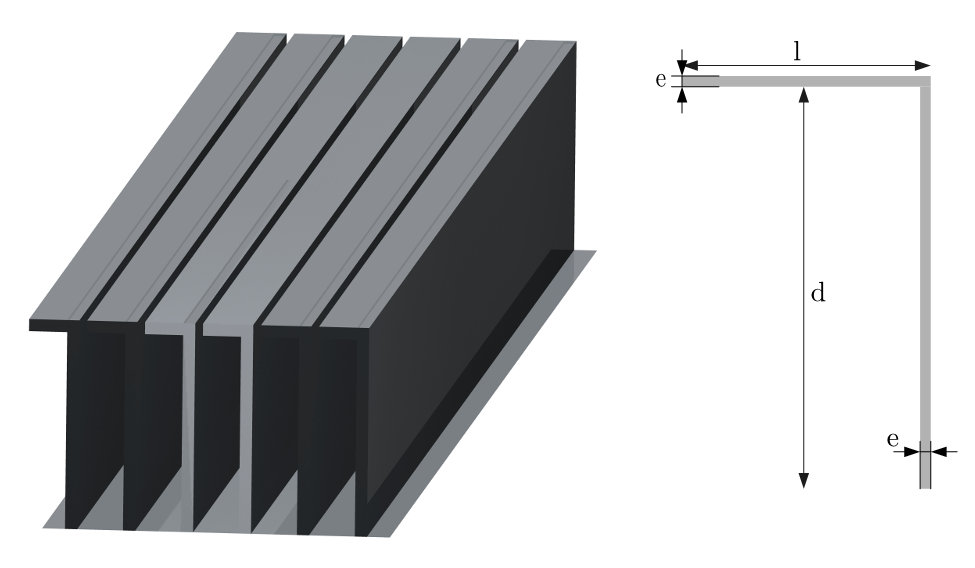}
		\centering 
		\caption{Inverted-``L'' in a simplified version of the original metamaterial where e=0.4, l=1.66 d=6.5[mm] and the free space between corrugations is 0.25[mm].}
		\label{fig3_6}
	\end{figure}

\begin{figure}[ht!]
		\includegraphics[width=0.6\textwidth]{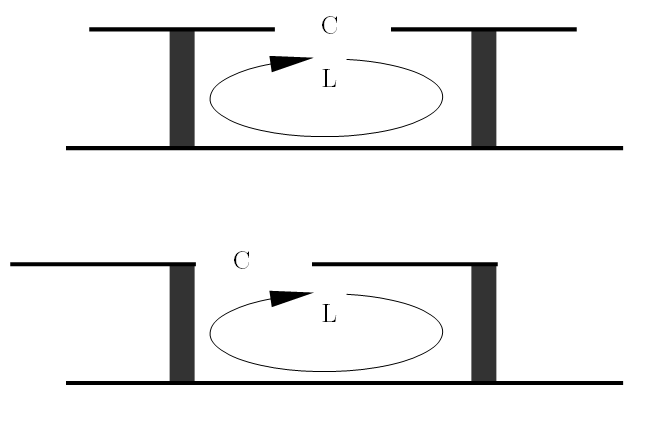}
		\centering 
		\caption{Illustration of the Top-plated (up) and inverted-``L'' (down) topologies similarity in terms of impedance. Since $R \simeq 0$ for good conductors, a generic impedance approximately adopts the value $Z=j\omega L-j/(\omega C)$ in both models, where $L$ and $C$ are given values of inductance and capacitance respectively. }
		\label{fig3_7}
	\end{figure}

Moreover, we reconverted our top-plated or ``T'' topology presented in fig.(\ref{fig3_5}) adapting it to a new thin version of the classic ring-loaded corrugations topology \cite{Clarricoats}, more frequently used in taper-converters for feedhorns. This thin-ring-loaded or inverted-``L'' topology is presented in Fig.\ref{fig3_6}. Simulations of this inverted-``L'' metamaterial indicate that both top-plated and inverted-``L'' topologies are equivalent from an electromagnetic point of view, so both could be combined or used for different applications depending on the constraints and requirements of each case. This electromagnetic similarity is logical because both typologies have identical surface inductance and capacitance. This idea can be qualitatively understood with the help of Fig.\ref{fig3_7}. The dotted black line in the Fig.\ref{fig3_4} which is named model-3 represents both top-plated and inverted-``L'' topologies. In this figure it can be seen that the results are better for this simplified model-3 than for model-1 or 2. This was initially surprising to the author. We think that this is due to several aspects of the design. Firstly, the reduction in the minimum thickness of the plates (adopting now the value 0.4[mm] for each thickness), but mainly because of the gain in radial symmetry in comparison with the metamaterial of models 1 and 2, where different points around the circumference can have different values of surface impedance due to the radial pattern. This has been confirmed by radially displacing the electric field probe in different simulations. 

Thus, it can be seen in fig. \ref{fig3_4}, that the operating band is theoretically even wider for the top-plated and inverted-``L'' metamaterial topologies (model-3), covering the 6.5-35[GHz] band for an impressive 5:1 bandwidth factor, and also because the value of $-Z^{S} Z^{P}/Z_0^2$ is closer to the unity over the whole band, theoretically presenting a better behavior in terms of cross polarization attenuation and satisfaction of the hybrid mode condition.Thus it can be concluded that, at least in theory, the metamaterial presented in \cite{ADA} does not have a real advantage over a simpler and more ``easily'' manufacturable thin version of the ring-loaded topology, presented in articles around 30 years ago by several authors (e.g.,\cite{Clarricoats}). Anyway, our interest in this article is to explore methods of fabrication  and more advantageous metamaterials for the future.

Finally, it is interesting to conclude this analysis comparing these new metamaterials with the more extended topology for the present-day design of feed-horns using corrugations. This reference helps in order to see how innovative these new metamaterials are. Thus, an equivalent version of a corrugated surface has been simulated in order to calculate its bandwidth. In this reference model, the depth of the slots is 6.9[mm], the thickness of the corrugations is 0.4[mm] and the lattice period or space between two consecutive lines of corrugations is an optimized value around 1.5[mm]. The result of the simulations of this model is represented by the dot-stripe gray line in Fig.\ref{fig3_4}, where it can be seen that the product of normalized impedances has a large slope so its bandwidth factor is much lower than the bandwidth factor of the metamaterial-based models, being around a 1.4:1 factor or 40\% of the band, as the bibliography predicts. \newline

The metamaterials presented in this article can satisfy the hybrid mode condition in wider bandwidths than that of the reference. This phenomenon can be explained qualitatively. The benefit in bandwidth with low levels of cross polarization of our meta-ring is based on two characteristics. Firstly, the gain in symmetry compared to those referenced geometries, i.e., the mushroom patterns are not symmetrical in their top plane. This lack of symmetry, caused by the teeth between consecutive mushroom plates, make the surface impedance to vary with the position in the upper plane of the structure so cannot be adjusted simultaneously through two consecutive zones, i.e., between plates and at the consecutive rectangular hollow space, yielding a global spurious effect and reducing the band. Secondly, in some cases, the reduction of thicknesses in comparison to the referenced mushroom geometries. The mechanical manufacturing in workshop allows to have much smaller thicknesses than those allowed by additive manufacturing or other techniques. This control over the thickness is transmitted to a control over the inductance and capacitance, which allow us to design a more precise structure in the fulfillment of the hybrid mode condition. On the other hand, regarding the classic corrugations, the bandwidth benefit of the metamaterials in the Fig.\ref{fig3_4} is produced due to the extra degree of freedom available to adjust the surface impedance values, given by the addition of top-plates.

\section{Material and methods}
\label{Methods}
\subsection{Materials justification}
Applications in feedhorn-antennas require the use of metals with high electric conductivity. Copper , titanium and aluminum were considered. 

A horn-antenna is a cylindrical-conical free-standing structure, so the only limitation in mechanical terms is to be able to manipulate it without deforming or breaking it, and that it be able to support cryogenic-vacuum conditions (to have adequate plasticity, resistance, and resilience).  Copper has the advantage of a higher conductivity, but aluminum is easier to mechanize. For each technique of fabrication presented we have taken the best material option which is always a compromise between electrical and mechanical properties. Anyway, both metals satisfies the basic conditions and they have been used in similar applications for our group for over twenty five years and in seven different telescopes and instruments (\cite{1994Natur.367..333H}, \cite{1998ApJ...498..117F},  \cite{2001MNRAS.327.1178G}, \cite{2003MNRAS.341.1057W}, \cite{JBO}, \cite{2014A&A...566A..54P}, \cite{2017MNRAS.464.4107G}). 

Furthermore, the penetration depth for both copper and aluminum at gigahertz frequencies is in the order of a few microns or lower. Because of this, we have considered the possibility of fabricating a metamaterial using SLA technology and covering with a metallic coating, because the microwaves do not the penetrate to the inner dielectric structure.

Table 1 compares the conductivity ($\sigma$) and penetration depth ($\delta$) of 3 pure metals (Al, Cu and Ti) at room temperature\footnote{Since $\sigma$ increase under cryogenics conditions, these characteristics are even more favorable at 4[K].} and 10[GHz] conditions \cite{Eskelinen}.

\begin{table}
\centering
\begin{tabular}{l*{3}{c}r}
\toprule Metal & $\sigma[S/m]$ & $\delta[\mu m]$ \\
\midrule
Cu               & $5.8\cdot10^7$ & 0.66 \\
Al            & $3.5\cdot10^7$ & 0.83 \\
Ti  & $2.5\cdot10^6$ & 3.7 \\

\bottomrule
\label{table_1}
\end{tabular} 
\caption{Conductivity ($\sigma$) and penetration depth ($\delta$) for Cu, Al and Ti at 300[K] and 10[GHz].}
\end{table} 

\subsection{Fabrication of advanced 3D metallic metamaterials}
 In this section we present two different ways of manufacturing the metamaterials presented in the section \ref{Theory}. We have focused our interest on two methods, i.e. Lost-wax printing and direct SLA 3D-printing \cite{Snyder}  being the most accurate techniques available and being the most promising for the geometries needed for  electromagnetic metamaterials at microwave frequencies.
 
 In the conceptual stage simulations were run on conical geometries with a diameter of 35[mm] and a height of 220[mm]. 
Due to the complexity of the geometry, 3D-printing presents itself as the most suitable method. 
Direct metal manufacturing was considered, and in particular Electron Beam Melting (EBM) \cite{Murr} in Ti6Al4v-eli. EBM does not require too much support structure, but  has not the resolution needed since the tolerances are of $\pm$0.2[mm]. Anyway, this method can be used with titanium which is not the most ideal since it has not the best conductivity (see table 1). 
 
  It is important to note that the use of a filling material instead of air was discarded early because in this case the metamaterial structure  must be scaled by a factor $\sqrt{\varepsilon_r}$, where $\varepsilon_r$ is the relative electric permittivity of the filling material. For example, in the case of polylactic acid or PLA, used commonly in 3D printing with $\varepsilon_r \sim 3$ at room temperature and 16[GHz] \cite{Dichtl}, the scale factor is around 1:1.7. That means all the geometrical dimensions of the metamaterial must be scaled by a factor 1:2 in order to operate in the same electromagnetic band, making the prototype, in principle, more difficult to manufacture. This solution will also complicate the use of the meta-feedhorn under cryogenic conditions, because of the differential contraction between metal and dielectric solids.
 
To conclude this section, we present a workshop mechanized version of the simplified metamaterial along with metrological studies of prototypes. 

\subsubsection{Metamaterial ring Lost-wax 3D-printing} 
\begin{figure}[ht!]
		\includegraphics[width=0.75\textwidth]{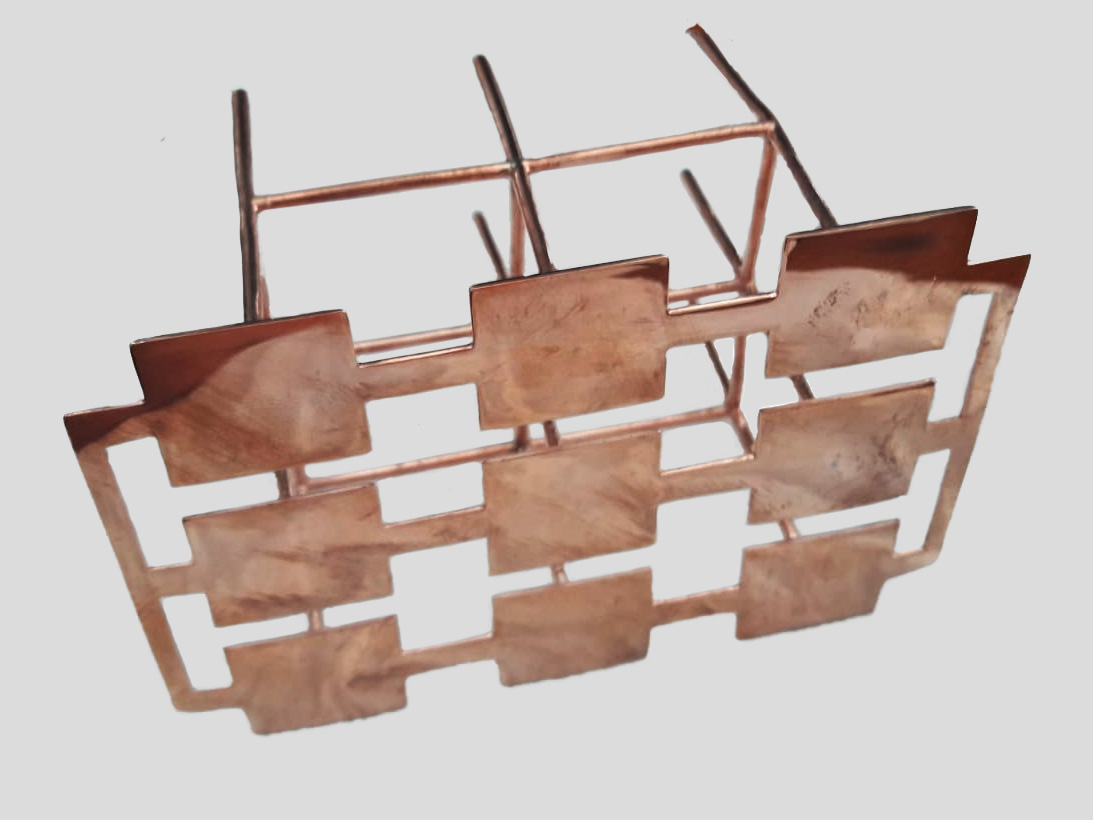}
		\centering 
		\caption{10:1 scaled version of the planar metamaterial of Fig.\ref{fig3_2}.}
		\label{fig3_8}
	\end{figure}

Firstly, we tried Lost-wax printing, where the 3D-model is built as a positive object through Stereo lithography, and later on prepared for traditional casting. The model and its support are attached to a wax “tree”, placed in a flask and covered in fine plaster. When the plaster solidifies, it forms a mold for the copper, silver, or gold casting. 
For SLA, generally a minimum thickness of 1[mm] is recommended and minimum features of 0.5[mm]. In the XY (in-plane) the material is fused with a 140$[\mu$m] laser spot and built up to a layer height of 50$[\mu$m]. This leads to adequate precision and the desired tolerances $\pm$0.1[mm] or better.
A first iteration was done at a 10:1 (i.e., ten times larger) scale, where we checked the tolerances on the scaled version as well as minimum limits. This prototype is shown in Fig.\ref{fig3_8}. To guarantee that the structure hold its shape reinforcements were added and the thickness adapted to manufacturing limits. 

\begin{figure}[ht!]
		\includegraphics[width=0.75\textwidth]{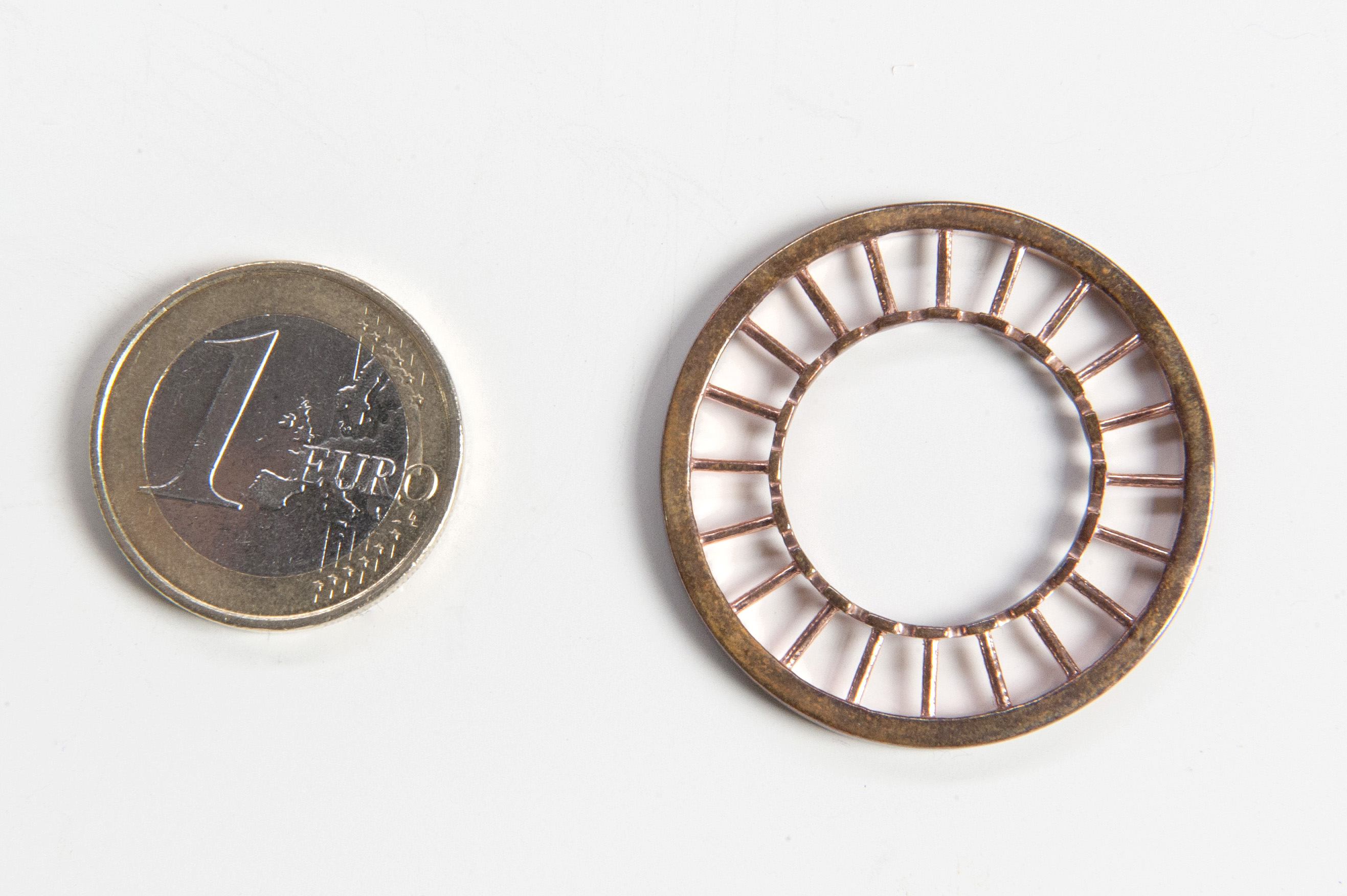}
		\centering 
		\caption{Ring structure metamaterial printed in copper by loss-wax technique.}
		\label{fig3_9}
	\end{figure}

Following this, one ring of a horn was printed by the same technique, presented here in Fig.\ref{fig3_9}. Dividing the cone structure into discs that are reassembled afterwards turns out to beneficial. Smaller, planar objects result in better geometrical tolerances, and more importantly require less support-structure, therefore less manual manipulation.  This is because most of the geometry rests on the build-plate. 

Deformation over large scales was significant. The nature of the process has a lot of manual operations: manipulating the mold, support removal, recovering the cast, polishing, packaging, shipping. Small dents and bumps lead to deformations, that end up giving unreliable and inconsistent results.  

\subsubsection{SLA prototyping of metamaterial periodic structures}

\begin{figure}[ht!]
		\includegraphics[width=0.5\textwidth]{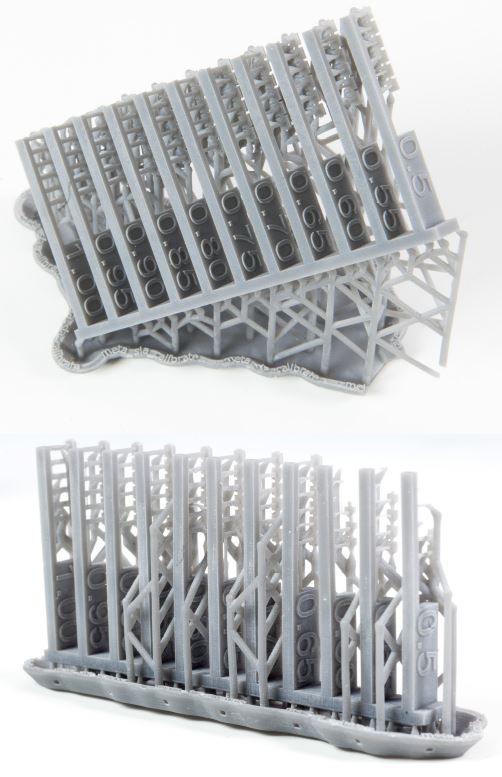}
		\centering 
		\caption{SLA resin iterations. It can be noted that the declination of the model (top)  is beneficial for printing dimensions which are forbidden for the planar version (bottom).}
		\label{fig3_11}
	\end{figure}

A second iteration was carried out in SLA resin printing where for both plaques and struts the thickness and the diameter was repeated from 0.5 to 1[mm] in 0.05 increments within the 3D structure in order to show feasibility of simulations which were run on geometries inferior to 1[mm] thickness. 
The results of the evaluation shown in Fig.\ref{fig3_11} reveal that  0.85[mm] and 0.9[mm] is consistently achievable. The viability of this technique is discussed in the following text.

The process of fabrication a film-substrate system generally generates stress in the mechanical pieces. Due to the differences in the coating and substrate properties, principally in the coefficient of thermal expansion (CTE), an additional thermal stress is generated on  cooling them. Different coating materials have been analyzed to find out feasibility of this process in our application. Despite the low conductivity of titanium alloys, these materials are evaluated as a starting point for this solution due to their mechanical properties. Ti-Al compound coatings are commonly applied in aerospace and automobile products. In fact, coatings obtained by Physical Vapor Disposition (PVD) techniques are widely applied to improve lifetime and performance  (e.g., \cite{Song}, \cite{Mo}).

In order to evaluate their behavior, it is necessary to make some simplifications that allow solving a parsimonious model. The simplified geometry to be evaluated consists of a dimensionally identical  ring of resin post-cured mixture of acrylate monomers, oligomers and photoinitiators with specific mechanical properties that are shown in table 2. The corners are eliminated avoiding stress concentrations. The base is covered with a metallic layer that is be prepared by a PVD technique. In addition, the study of different films is made by choosing different materials to evaluate the dependence of the mechanical properties of the film.

In this section, the structural interaction of the resin with a coating of 50 microns after decreasing the temperature to reach cryogenic conditions is analyzed using finite element method, more specifically the ANSYS Workbench$^{\textregistered}$. The results were correlated using reduced equations of the analytical solution in the case of thin-walled concentric cylinders to validate the results of the computational model.
In the analysis, the contact condition between components is rigid. It means that there is no slip in the surface region and the residual stress is not taken into account in the calculation. On the other hand, the relation of stiffness-thickness of the coating respect to the substrate is not significant. As a consequence of the fact that the thermal deformation is clearly more significant in the substrate ($\Delta r_s >> \Delta r_c$), we can assume that the coating does not provide any rigidity to the assembly and the deformations are mainly due to the cooling of the resin. Indeed, it can be considered that  the thermal displacements of the base are the sole course of the tensional state.

\begin{table}[]
\centering
\begin{tabular}{@{}ccccc@{}}
\toprule
Material   & UTS [MPa] & TM [GPa] & Elongation [\%] & TE [$^\circ C^{-1}$] \\ \midrule
Green      & 35                              & 1.4                   & 32.5            &    -               \\
Post-Cured & 61                              & 2.6                   & 13              & 7.85E-5                  \\ \bottomrule
\end{tabular}
\label{table_2}
\caption{Resin properties, where Ultimate Tensile Strength is UTS, Tensile Modulus is TM and TE is Thermal Expansion.}
\end{table}

\begin{table}[]
\begin{tabular}{@{}cccc@{}}
\toprule
Configuration      & Type      & Equiv. Lineal Strain [\%] & Equiv. Lineal Stress [MPa] \\ \midrule
Resin/Copper Alloy & Substrate & 0.0268                        & 68.86                          \\
Resin/Copper Alloy & Coating   & 0.0100                        & 1101.4                         \\
Resin/Ti6Al4V      & Substrate & 0.0276                        & 71.65                          \\
Resin/Ti6Al4V      & Coating   & 0.0063                        & 707.36                         \\
Resin/AA6061-T6    & Substrate & 0.0268                        & 69.38                          \\
Resin/ AA6061-T6   & Coating   & 0.0098                        & 684.12                         \\ \bottomrule

\label{table_3}

\end{tabular}
\caption{Stress State of film-Substrate.}
\end{table}

The results for the different materials in a cryogenic cool-down process are shown in table 3. From the results it is verified that the deformations caused by the variation of the temperature are practically identical with and without the coating, $\varepsilon \simeq 0.027$. The values obtained for the resin in each case are similar. As has been mentioned previously, the thermal contraction and the stiffness can be assumed null because the coating thickness is small when compared with the substrate.
The titanium alloy is the only material capable of withstanding the requirements in its elastic state. However, only one cold cycle has been evaluated, so the fatigue conditions and the concentration of tensions in the geometry of the antenna will eventually crack the coating. Furthermore, the rigidity of titanium is much higher than resin. The coating loads the substrate reaching its state of load-limit for just one temperature  cycle, which is why it would not stand a reasonable life span which consists of many cooldowns. In conclusion, it is not feasible to use this solution for this application without previously testing and proving the assembly.

\subsubsection{Mechanization of a simplified version of the metamaterial} 

\begin{figure}[ht!]
		\includegraphics[width=0.75\textwidth]{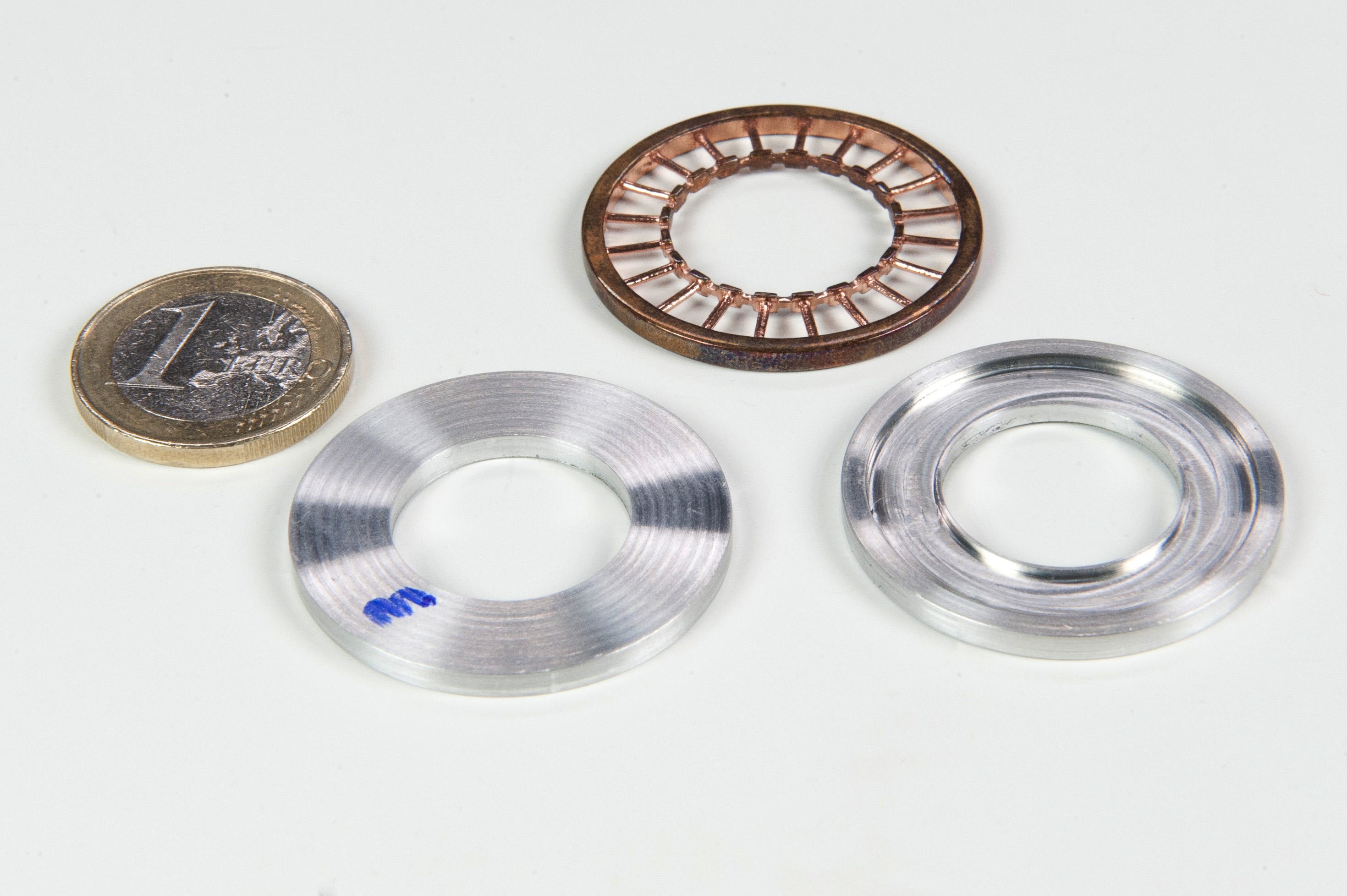}
		\centering 
		\caption{Simplified workshop mechanized prototype based on the model in Fig.\ref{fig3_6}.}
		\label{fig3_12}
	\end{figure}

Two simplified prototypes of the meta-horn corrugation were manufactured by traditional machining processes, milling and turning, in order to compare their capabilities to the meta-material ones. Both rings are similar, they have the same wall thickness while the corrugation profile is different (see Fig.\ref{fig3_12}).

The lathe and milling machine must apply a certain force to cut the material, so the minimum thickness that can be achieved is related to the material stiffness. Therefore, aluminium AW-6061-T6 was chosen instead of copper because it has better machinability. 
Due to the thickness and the required tolerances, of the order of 0.4[mm] and 0.050[mm] respectively, some supporting tools were made prior to manufacturing in order to control the deformation during the process.
The examples presented are only a prototype for proving the capability of milling and turning when sub-millimetre thickness is required. Hence, pieces were made as rings without any assembly interface. Since the antenna will be made from several disks assembled together, an interface must be designed for this purpose, which supposes a challenging mechanical work in itself as it must ensure the concentricity of each disk with respect to any other for a good feedhorn performance. Anyway, practical ring junctions are well established and can be done, for example, with the technique published in \cite{Milano}.

Metrological results have been collected for all the manufactured metamaterials and are presented in section \ref{Results}.

\section{Results}
\label{Results}

\begin{figure}[ht!]
		\includegraphics[width=0.9\textwidth]{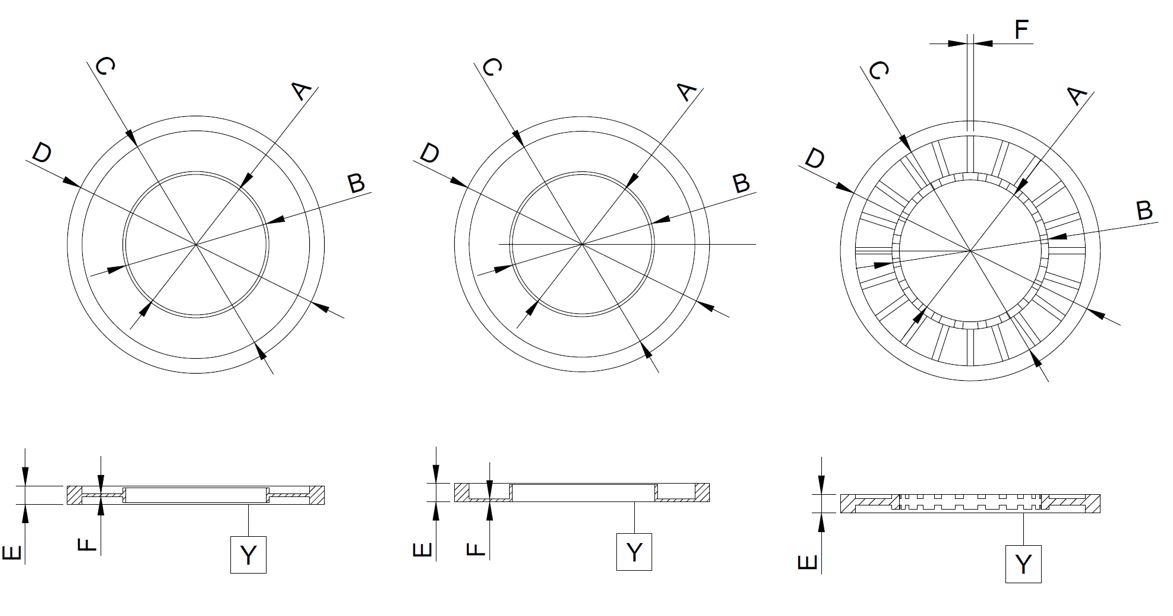}
		\centering 
		\caption{Schematic of top-plated corrugations of model-3 (left), inverted-``L'' typology of model-3 (middle) and mushroom type metamaterial of model-2 (right). Note the dimensions presented in table \ref{table_4}}.
		\label{fig3_13}
	\end{figure}

In order to evaluate the acceptability of the different manufacturing processes, a metrology analysis was performed for each piece. In this analysis, both dimensional and geometric tolerances, i.e., roundness, parallelism and flatness, are measured. A schematic drawing of the pieces is shown in Fig.\ref{fig3_13}. The results are summarized in table \ref{table_4}.

\begin{table}[h]
\begin{tabular}{ccccccc} \toprule 
            &         & \multicolumn{3}{c}{Model-3}                & \multicolumn{2}{c}{Model-2} \\
            &         &              & Top-plated   & Inverted-``L''  &              &              \\
Dimension   & Measure & Nom. {[}mm{]} & Dev. {[}$\mu$m{]} & Dev. {[}$\mu$m{]} & Nom.{[}mm{]}  & Dev.{[}$\mu$m{]}  \\ \midrule
A (inner D) & D       & 19.000       & -2           & -7           & 19.000       & +91           \\
            & R       &              & 0            & +3            &              & +100          \\
B           & D       & 19.800       & -30          & -15          & 21.000       & -262         \\
            & R       &              & +3            & +2            &              & +127          \\
C           & D       & 30.800       & +28           & +11           & 30.800       & -273         \\
            & R       &              & +3            & +7            &              & +42           \\
D (outer D) & D       & 34.800       & +15           & -13          & 34.800       & -407         \\
            & R       &              & +12           & +15           &              & +91           \\
E           & T       & 2.500        & -30          & -23          & 2.500        &              \\
            & P       &              & +14           & +10           &              &              \\
            & F       &              & +8            & +9            &              &              \\
F           & T       & 0.400        & -48          & -8           & 0.900        & +50           \\
            & P       &              & +7            & +20           & R            & +12           \\
            & F       &              & +16           & +12           &              &             

\end{tabular}
%\bottomrule
\caption{Summary of metrology where the nominal (nom.) and deviation (dev.) in the dimensions highlited in the Fig.\ref{fig3_13} are taken and the measured properties are D=diameter, R=Roundness, T=thickness, P=parallelism referred to ``Y'' and F=flatness.}
\label{table_4}

\end{table}

Table 4 shows how the dimensions of the aluminum pieces manufactured by turning/milling have a deviation less than 0,050[mm] from the nominal values. For the meta-horn prototype manufactured by additive printing the deviation is higher and exceeds the acceptable values for a good feedhorn performance, established around 100[$\mu m$] for frequencies of tens of gigahertz. 

\section{Discussion and Conclusions}
\label{Conclusions}

The key to the prototyping of metamaterials is the small high precision parts. 3D-printing processes require support structures and these can be difficult or impossible to remove without damaging the model. 
Strategic design can drastically improve the situation and reduce the need for supports, which leads to better 3D-printed results, and tolerances. We have found that the lost-wax method doesn't seem to allow us to fabricate metamaterials at microwave frequencies with the required precision, and that the SLA+metallic-skin method suffers from breakage of the metamaterial in multiple cryogenic cycles. Workshop machining has been presented as a reliable option.\newline

\begin{figure}[ht!]
		\includegraphics[width=1\textwidth]{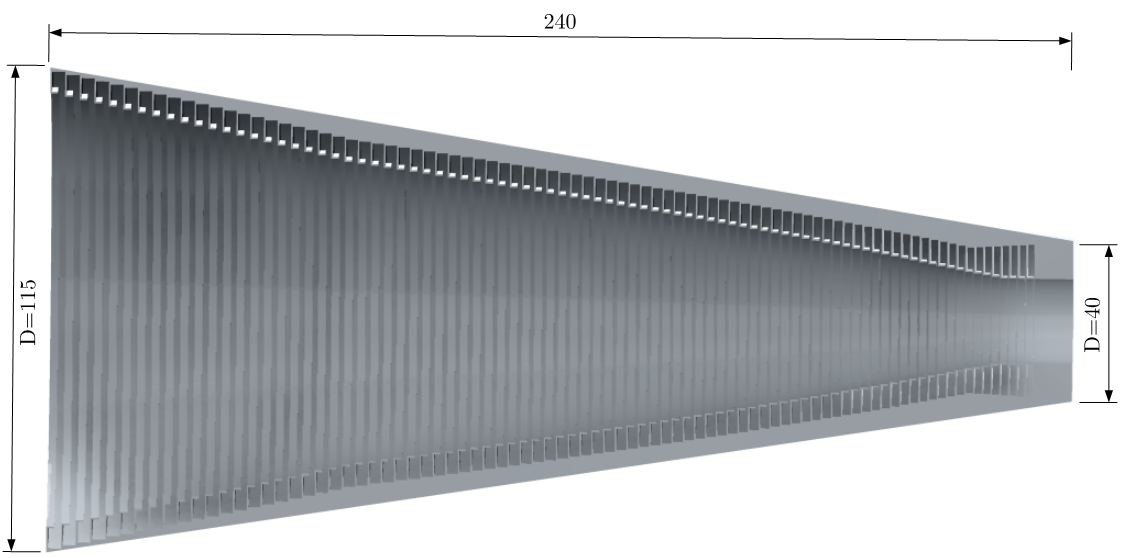}
		\centering 
		\caption{Design of a meta-horn. General dimensions, in millimeters.}
		\label{fig_14}
	\end{figure}
\begin{figure}[ht!]
		\includegraphics[width=0.8\textwidth]{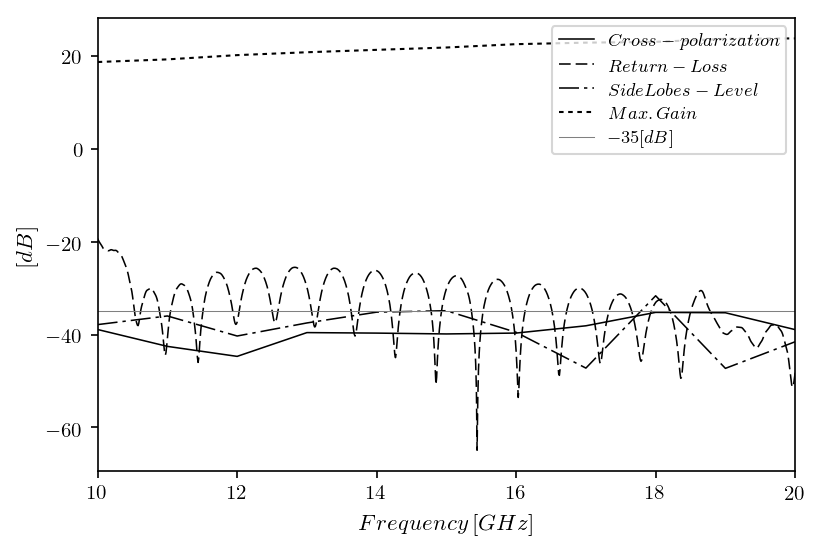}
		\centering 
		\caption{Design of a meta-horn. Results of the simulations.}
		\label{fig_15}
	\end{figure}
\begin{figure}[ht!]
\centering
		\includegraphics[width=1\textwidth]{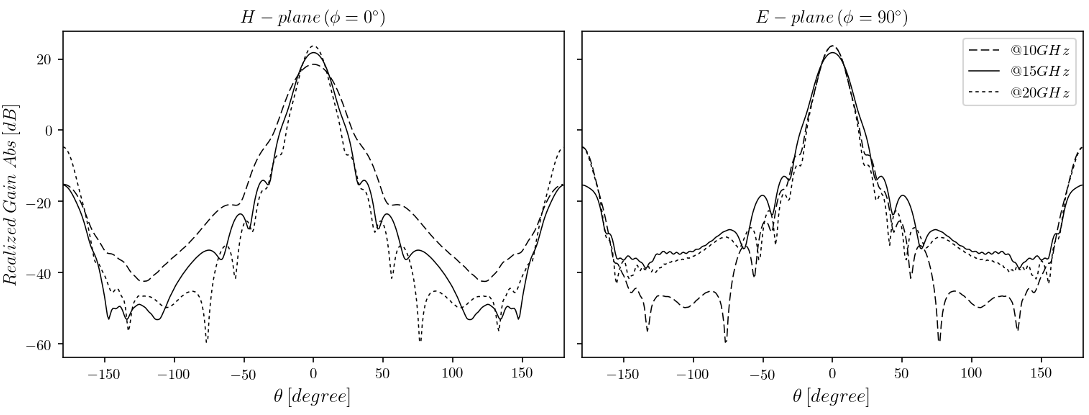}
		\centering 
		\caption{E and H plane farfield antenna patterns at 10, 15 and 20[GHz].}
		\label{fig_16}
		\vspace{-\baselineskip}% remove one line of space below this figure caption
	\end{figure}
\begin{figure}[ht!]
\centering
		\includegraphics[width=1\textwidth]{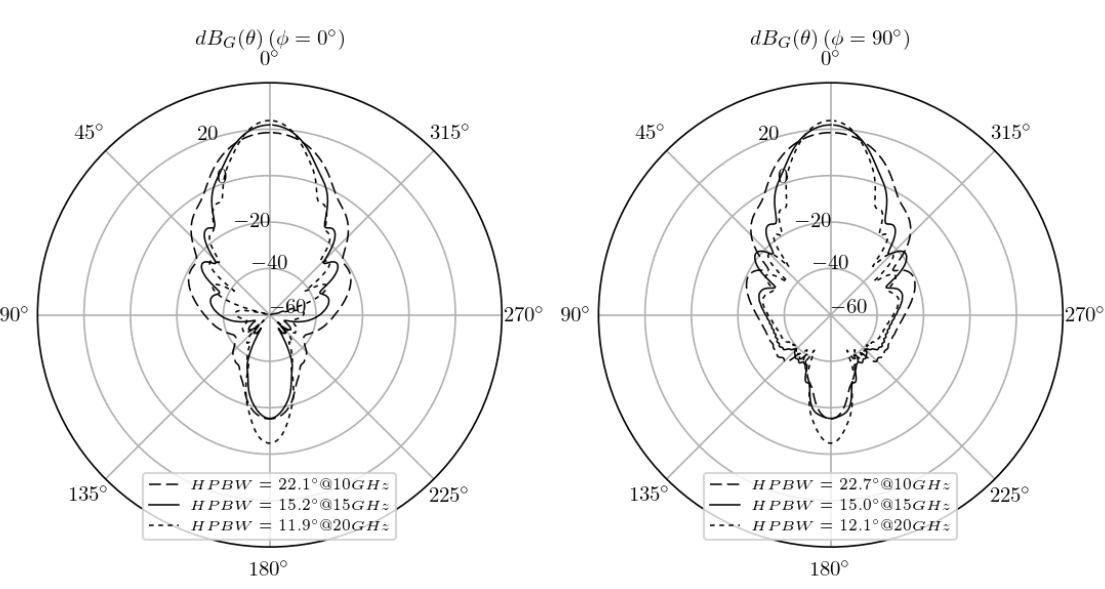}
		\centering 
		\caption{Farfield realized gain patterns in decibels ($dB_G$) at 10, 15 and 20[GHz] for $\phi=0^{\circ}$ and $\phi=90^{\circ}$ planes. Angular width at -3[dB] (HPBW) given for each case.}
		\label{fig_17}
		\vspace{-\baselineskip}% remove one line of space below this figure caption
	\end{figure}

Here we show a more practical and usable example using the meta-ring of Fig.\ref{fig3_12} to design a conical feedhorn composed of meta-rings. The design of the horn and the very promising results based on CST Studio Suite$^{\textregistered}$ simulations are shown in Figs.\ref{fig_14} and \ref{fig_15}. In Fig.\ref{fig_15} an excellent behaviour over a entire 2:1 bandwidth in terms of return-loss, side-lobe levels, gain and particularly cross-polarization, where levels better than -35[dB] over the entire 10-20[GHz] band are predicted. 
The E and H planes demonstration is shown in the Fig.\ref{fig_16}, where an excellent behaviour is presented at the central frequency and the limits of the band of work. The polar farfield realized gain pattern diagrams are shown in Fig.\ref{fig_17} for both $\phi=90^{\circ}$ and $\phi=0^{\circ}$ planes, where the \textit {half power beam width} (HPBW) presents exceptional results as well.
\newline

This design is very compact even using a simple conical profile because an easily manufacturing solution was required, but the characteristics could be improved further by using a more adequate profile \cite{Miguel_Hern_ndez_2019}.

%% References
%%
%% Following citation commands can be used in the body text:
%% Usage of \cite is as follows:
%%   \cite{key}         ==>>  [#]
%%   \cite[chap. 2]{key} ==>> [#, chap. 2]
%%

%% References with bibTeX database:

\bibliographystyle{elsarticle-num}

\bibliography{sample}

\end{document}